\documentclass{emulateapj}

\usepackage{psfig}	%%%the original one
\usepackage{natbib}
\shorttitle{Far-Ultraviolet Observations of IC 405}
\shortauthors{France et al.}

\begin{document}


\title{Rocket and $FUSE$ Observations of IC 405: \\
    Differential Extinction and	Fluorescent Molecular Hydrogen}


\author{K. France}%\altaffilmark{1}}
\affil{Department of Physics and Astronomy, Johns Hopkins University,
    Baltimore, MD 21218}

\author{S. R. McCandliss}%\altaffilmark{1}}
\affil{Department of Physics and Astronomy, Johns Hopkins University,
    Baltimore, MD 21218}

\author{E. B. Burgh}%\altaffilmark{2}}
\affil{Space Astronomy Laboratory, University of Wisconsin,
    Madison, WI 53706}
    
\and

\author{P. D. Feldman}%\altaffilmark{1}}
\affil{Department of Physics and Astronomy, Johns Hopkins University,
    Baltimore, MD 21218}
    

%%%These are for extra affiliations at the bottom of page
%\altaffiltext{1}{Visiting Astronomer, Cerro Tololo Inter-American Observatory.
%CTIO is operated by AURA, Inc.\ under contract to the National Science
%\altaffiltext{2}{present address: Center for Astrophysics,
%    60 Garden Street, Cambridge, MA 02138}
%\altaffiltext{3}{Patron, Alonso's Bar and Grill}


\begin{abstract}
We present far-ultraviolet spectroscopy of the emission/reflection nebula IC 405 
obtained by a rocket-borne long-slit spectrograph and the 
$Far$~$Ultraviolet$~$Spectroscopic$~$Explorer$.  
Both data sets show a rise in the ratio of the nebular 
surface brightness to stellar flux (S/F$_{\star}$) of approximately 
two orders of magnitude 
towards the blue end of the far-UV bandpass.   
Scattering models using simple dust geometries fail to reproduce the 
observed S/F$_{\star}$ for realistic grain properties. 
The high spectral resolution of the $FUSE$ data 
reveals a rich fluorescent molecular hydrogen 
spectrum $\approx$~1000\arcsec~north of the star that is clearly
distinguished from the steady blue continuum.  
The S/F$_{\star}$ remains roughly constant at all nebular pointings, 
showing that fluorescent molecular hydrogen is not the dominant
cause for the blue rise.
We discuss three possible mechanisms for the ``Blue Dust'': differential
extinction of the dominant star (HD~34078), 
unusual dust grain properties, and emission from nebular dust.  
We conclude that
uncertainties in the nebular geometry and the degree of dust clumping are 
most likely responsible for the blue rise.
As an interesting consequence 
of this result, we consider how IC 405 would appear in a spatially unresolved observation.
If IC 405 was observed with a spatial resolution of less than 0.4 pc, for example,     
an observer would infer a far-UV flux that was 2.5 times the true value, giving the 
appearance of a stellar continuum that was less extinguished than radiation 
from the surrounding nebula, an effect that is reminiscent of the observed ultraviolet
properties of starburst galaxies.
\end{abstract}



%%%%%%%%%%%%%%%%%%%%%%%%%%%START OF THE PAPER%%%%%%%%%%%%%%%%%%%%%%%
\section{Introduction}

Observations aimed at determining the far-ultraviolet properties of dust grains are 
challenging as they require space-borne instrumentation designed 
specifically for such studies.  The lack of data has hampered testing
models of interstellar dust grains in the far-ultraviolet
(far-UV) spectral region below Lyman-$\alpha$ where dust most strongly
attenuates the radiation field.  Extinction by dust grains plays a key role
in regulating the escape of radiation from a hot-star environment,
thereby determining the physical structure and chemical
equilibrium of the local interstellar medium (ISM). 
Constraints can be placed on the physical properties of dust grains 
by studying how they absorb and scatter
far-UV light where the two interact, either near a bright
source of ultraviolet photons or in the diffuse interstellar medium.  
Reflection nebulae offer an environment where a strong far-UV radiation
field is interacting with a dusty, gaseous environment.
This class of objects has been used by a number of observers 
in an attempt to derive the 
properties of grains (see Draine 2003a for a recent review).~\nocite{draine03ar}
The interpretation of these observations is complicated by geometric uncertainties 
in the nebular dust distribution, contamination by emission
from atomic and molecular species within the nebula, and  
uncertainties in the incident spectral energy distribution~\citep{witt93,ngc2023,hurwitz98}.
Burgh et al. (2002) have made the first measurement 
of the dust albedo ($a$) and phase function asymmetry parameter 
($g$~$\equiv$~$\langle$cos~$\theta\rangle$) below Ly-$\alpha$,
using long-slit spectroscopy to constrain both quantities.
However,~\citet{mathis02} have argued that observations of reflection
nebulae are so confused by the clumpy nature of the dust distributions that 
the derived grain properties are highly unreliable.
Further addressing this problem, we present far-UV observations
of IC 405.  

\begin{figure}
\begin{center}
\epsscale{1.0}
\rotatebox{0}{
\plotone{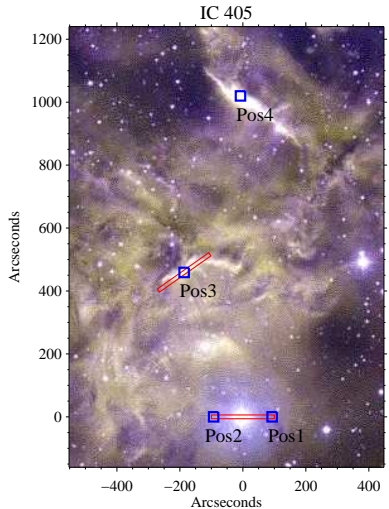} }
\caption{\label{slits}IC 405 with the relevant aperture overlays, HD 34078 is 
located at the origin.  The long slits represent 
the aperture of the rocket spectrograph, and the boxes represent the $FUSE$ LWRS 
aperture. Optical image from $APOD$, 
T.A.Rector and B.A.Wolpa/NOAO/AURA/NSF.}
\end{center}
\end{figure}

IC 405 (The Flaming Star Nebula) is an  emission/reflection~\citep{herbig99} nebula in Auriga, 
illuminated by a dominant star, AE Aur (HD~34078).  
AE Aur is one of the stars thought to have been ejected from 
the Orion region roughly 2.5 million years 
ago in a binary-binary interaction that led to the creation of the well 
studied $\iota$ Ori binary system~\citep{runaway}.  Consequently, AE Aur 
is moving with a large proper motion through the nebula at
$\approx$~40 milliarcseconds yr$^{-1}$~\citep{405v_tan}.
Herbig (1999) used repeated observations of this
high proper motion object to search for interstellar line variations, providing 
information about clump structure in the ISM~\nocite{herbig99}. 
Presently, AE Aur is thought to be cospatial with the nebula at a distance of about 
450 pc.  It is bright in both the visible (V~$=$~6.0) and the ultraviolet 
(O9.5~Ve), although it is rather extinguished 
($E(B~-~V)~=~0.53$ and $R_{V}$~=~3.42; Cardelli, Clayton, $\&$ Mathis 1989).
In this paper, we present far-UV 
observations of IC 405 made with a rocket-borne long-slit spectrograph
and the $Far$~$Ultraviolet$~$Spectroscopic$~$Explorer$ ($FUSE$).  
Figure~\ref{slits} shows the nebula with the various spectrograph apertures 
overlaid.  Section 2 includes a description of the rocket experiment, 
the observations, and the data reduction.  In $\S$~3, 
we describe the $FUSE$ observations and 
reduction.  Section 4 puts the IC 405 data in context with 
previous studies of other reflection nebulae, including
discussion and interpretation of the results and their possible consequences.  
Our observations of IC 405 are summarized in~$\S$~5.

\section{Rocket Observations}

Long-slit spectroscopy is a technique that lends itself to the study 
of reflection nebulae where the exciting star(s) typically 
are embedded in or near the gas and dust with which they 
are interacting.   The extended aperture allows nebular spectra to 
be measured continuously along the slit, and variations can be observed
in both wavelength and angle from the central object.  This
information can also be obtained through small apertures, but
this requires multiple observations, which is costly in the far-UV bandpass where
observing time is at a premium.

\subsection{Instrument Description and Observations}

The sounding rocket experiment consists of a telescope and spectrograph designed 
for use in the far-UV.  The telescope is an updated version
of the Faint Object Telescope (FOT), 
a 40 cm diameter, f/15.7 Dall-Kirkham~\citep{fot,srm94,ngc2023}.  
The optics are coated with a layer of ion-beam
sputtered SiC to enhance the reflectivity at far-UV wavelengths.  The telescope
is housed in an invar heat-shield with a co-axially mounted startracker
which provides error signals to the Attitude Control System (ACS).

\begin{figure*}
\begin{center}
\epsscale{0.5}
\rotatebox{90}{
\plotone{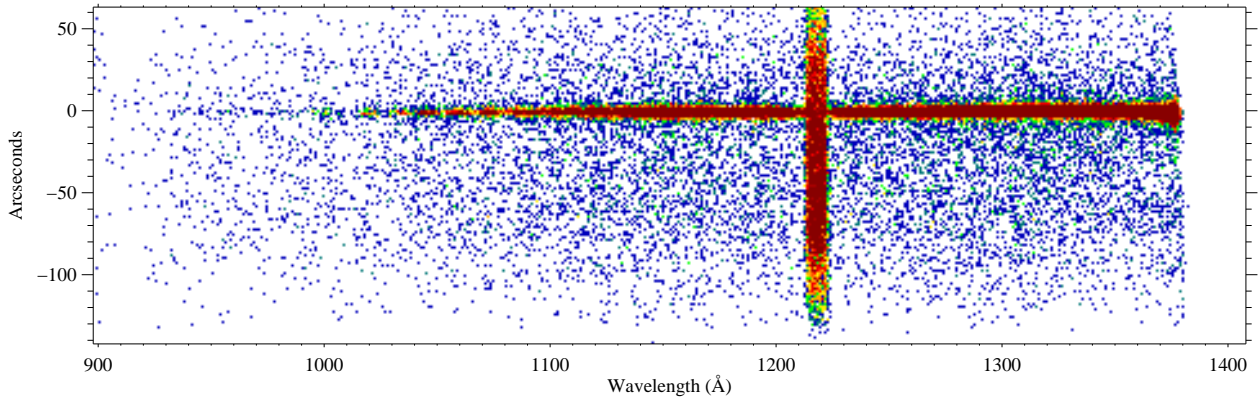}}
\caption{\label{flightdata}The spectrogram of the rocket data, following  
time-tagged correction for pointing.  HD 34078 is the horizontal
strip, the nebular spectra are above and below the star, and
the broad vertical feature is geocoronal Ly-$\alpha$.}
\end{center}
\end{figure*}

The FOT focuses the target at the entrance aperture of the instrument, 
an evacuated 400~mm diameter Rowland Circle spectrograph 
using a holographically ruled, SiC coated, diffraction grating in the first 
order.  The detector is a microchannel plate stack with a 
KBr photocathode, read out by a double delay-line anode~\citep{ossy93}.  The spectrograph 
is kept at a vacuum of $\approx$ 
10$^{-8}$~Torr and isolated from the spectrograph section by a gate valve that
opens under timer control in flight.  The spectrograph and telescope 
sections share a common vacuum ($<$~few~$\times$~10$^{-5}$~Torr).  
A mirrored slitjaw, angled 22$^\circ$ from the optical axis, 
lies at the telescope focus.   A long slit 
(12\arcsec$\times$200\arcsec\ projected on the sky, 5.6~$\times$~10$^{-8}$~sr) is etched into the slitjaw,
defining the entrance aperture to the spectrograph.  The spectrograph achieves a 
pointing limited spectral resolution of $\approx$~3~\AA.

This experiment was launched aboard a Black Brant IX
sounding rocket (NASA 36.198~UG) from White Sands Missile
Range, New Mexico ($106\fdg3$ West, $32\fdg4$ North), on 09 February 
2001 at 21:00 MST.  The target was obtained by referencing the startracker 
to two bright guide stars (Sirius and Capella), then reorienting to the 
target.  The acquired field is within a few arcminutes of the expected pointing.
The field is relayed to the ground in real-time through a Xybion TV 
camera imaging the slitjaw (20\arcmin~field-of-view).  Fine adjustments are 
performed in real-time via commands to the ACS.
HD 34078 was placed in the spectrograph slit at T+150 seconds and data was 
taken continuously until detector high-voltage turn-off at T+490 seconds.
The stellar spectrum was recorded for 106.2 seconds of the flight 
(hereafter the ``on-star'' pointing).  Two offset pointings within the nebula were 
observed with only one having a count rate higher than the background 
(observed for 68.7 seconds, hereafter the ``off-star'' pointing).  
Figure~\ref{flightdata} shows the pointing corrected long-slit spectrogram of the 
data obtained from target acquisition to detector turn-off.

\subsection{Analysis of the Rocket Observations}

Flight data were analyzed using IDL code customized to read the data as
supplied by the telemetry system.  A background subtraction can be made by 
measuring the flux on the detector after instrument turn-on, but prior to 
target acquisition.  The primary source of background flux is geocoronal 
airglow from atomic hydrogen and oxygen.  The data are then calibrated with 
measurements of the telescope mirror reflectivities and spectrograph
quantum efficiency, measured both before and after flight in the calibration 
facilities located at The Johns Hopkins University.

\begin{figure}[b]
\begin{center}
\epsscale{0.45}
\rotatebox{90}{
\plotone{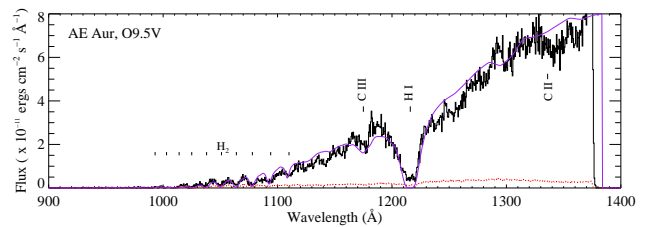} }
\caption{\label{histspec}The spectrum of HD~34078, overplotted 
with a stellar model extinguished by the parametrization of Fitzpatrick
and Massa and a model of H$_{2}$ absorption.}
\end{center}
\end{figure}

\begin{figure}[b]
\begin{center}
\epsscale{0.88}
\rotatebox{90}{
\plotone{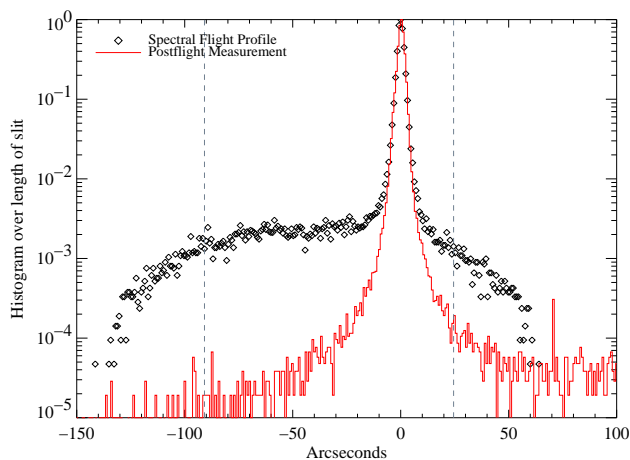} }
\caption{\label{lsf}Spatial profile of the 
flight data in black (excluding Ly-$\alpha$ airglow). 
Postflight determination of the instrument line-spread-function 
reproduces flight profile.  The dashed lines represent the portion of 
the spectrograph slit unaffected by instrumental vignetting.
One notices the extension of the nebula  beyond the stellar peak.}
\end{center}
\end{figure}

\begin{figure}[b]
\begin{center}
\epsscale{0.48}
\rotatebox{90}{
\plotone{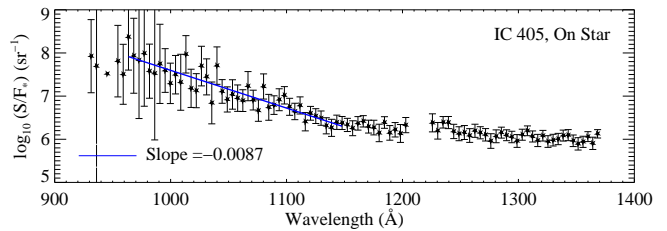}}
\caption{\label{sovf_rckt}The ratio of nebular surface 
brightness to stellar flux measured by the sounding rocket 
experiment.  Note the rise of approximately two orders of 
magnitude to the blue across the bandpass.}
\end{center}
\end{figure}

The stellar spectrum is extracted following corrections for pointing errors
and detector drift.
Figure~\ref{histspec} shows the spectrum of the HD~34078 obtained during 
the flight.  The spectrum is of high quality (S/N $\approx$ 10-15 at $R$~=~300),
and is consistent with previous far-UV measurements 
($IUE$; Penny et al. 1996 and $FUSE$; Le Petite et al. 2001).\nocite{penny96,fuse2}  In Figure~\ref{histspec}, the 
measured spectrum has been shown overplotted with a synthetic
stellar spectrum.  The model has been extinguished using
the parametrization of~\citet{fm90} and the $H_{2}ools$ molecular hydrogen
absorption templates~\citep{h2ools} for a $b$~=~3, as determined by Le Petite et al. (2001),
and a total column density $N$(H$_{2}$)~=~8.1~x~10$^{20}$~cm$^{-2}$, that we
determined from the $FUSE$ spectrum of the star, described below.
The stellar spectrum shows
interstellar absorption features of \ion{H}{1}, H$_{2}$, \ion{C}{2}, and 
\ion{O}{1} and photospheric \ion{C}{3} absorption.  
After the star is removed, the spectra from different regions within
the nebula can be separated by ``time-tagging'' the data from a playback
of the slitjaw camera that shows the flight time.  Integrating the nebular 
spectra over the area of the slit allows one
to measure the nebular surface brightness.  
Using the
vacuum collimator described by~\citet{vaccoll}, we can determine the 
instrumental line-spread-function, and we find that the in-flight 
stellar profile matches our post-flight calibrations.
Figure~\ref{lsf} shows the spatial profile of the nebular brightness (corrected
for Ly-$\alpha$ airglow) that extends well beyond the intrinsic instrumental profile.  
Once the nebular surface brightness 
and stellar flux have been measured, their ratio (S/F$_{\star}$) 
can be taken.
This reveals the most surprising feature of our observations: 
the ratio of nebular surface brightness to stellar flux rises 
roughly two orders of magnitude to the blue from 1400 to 900 \AA, as 
shown in Figure~\ref{sovf_rckt}.  This 
sharp rise in S/F$_{\star}$ is in stark contrast with similar observations
of the reflection nebulae NGC 2023 and 7023~\citep{ngc2023,witt93,murthy93}, where 
the S/F$_{\star}$ is found to be constant with wavelength in the far-UV regime.

\section{$FUSE$ Observations and Data Reduction}

Additional nebular observations were made by $FUSE$ from 11 to 13 March 2003 
on four positions within IC 405 (programs D12701, 02, 03, and 04).  
Spectra were obtained in the 905--1187~\AA\ bandpass with the 
low-resolution (LWRS) aperture (30\arcsec$\times$30\arcsec) on $FUSE$ 
(see Moos et al. 2000 for a satellite description and Sanhow et al. 2000
for on-orbit performance characteristics).~\nocite{moos00,sanhow00} 
Assuming that the nebulosity fills the LWRS aperture, a spectral resolution 
of 0.33 \AA\ is achieved at 1060 \AA.
Positions 01 (Pos1) and 02 (Pos2) were observed for 1850 and 1880 seconds, 
respectively, corresponding to 
the bottom and top of the long-slit of the rocket spectrograph.  
Position 03 (Pos3) was observed for 12.6 ks, overlapping with the 
``off-star'' rocket pointing and Position 4 (Pos4), 
a bright optical filament north of the star was observed for 
12.4 ks. 
Using an approach similar to the unsharp masking technique
described by~\citet{unsharp}, we suspect Pos4 
to exhibit Extended Red Emission (ERE).  A list of the $FUSE$ 
pointings is given in Table~\ref{fuse_sum}.

\begin{deluxetable*}{cccccc}
\tabletypesize{\Small}
\tablecaption{Summary of $FUSE$ observations of IC 405. \label{fuse_sum}}
\tablewidth{0pt}
\tablehead{
\colhead{Position} & \colhead{Program}   & \colhead{RA (2000)}   &
\colhead{$\delta$ (2000)} & \colhead{Integration Time}  & 
\colhead{Brightness at 1060 \AA} \\ 
 &   &   \colhead{( $^{\mathrm h}$\, $^{\mathrm m}$\, $^{\mathrm s} )$} &
\colhead{(\arcdeg\, \arcmin\, \arcsec )}     & \colhead{(seconds)}  &
\colhead{(ergs cm$^{-2}$ s$^{-1}$ \AA$^{-1}$ sr$^{-1}$)}
}
%\tableline
\startdata
IC405-01 & D12701 & 05 16 12.15 & +34 18 49.3 & 1850 &  2.0 x 10$^{-5}$ \\
IC405-02 & D12701 & 05 16 24.15 & +34 18 49.3 & 1880 &  2.7 x 10$^{-5}$ \\
IC405-03 & D12703 & 05 16 30.21 & +34 24 56.4 & 12630 & 3.5 x 10$^{-6}$ \\
IC405-04 & D12704 & 05 16 18.64 & +34 32 25.2 & 12445 & 1.1 x 10$^{-6}$ \\
 \enddata

\end{deluxetable*}

Data for all four pointings were obtained in ``time-tagged'' (TTAG) mode
and were initially processed using the CALFUSE pipeline, version 2.2.3. 
The calibrated data files for each orbit were then combined using IDL 
software, and when possible, the individual channels were combined using 
a cross-correlation algorithm.  Due to the diffuse nature of the targets, 
there were instances where this algorithm failed; in these cases, 
the channels were analyzed independently.

\begin{figure}
\begin{center}
\epsscale{0.9}
\rotatebox{90}{
\plotone{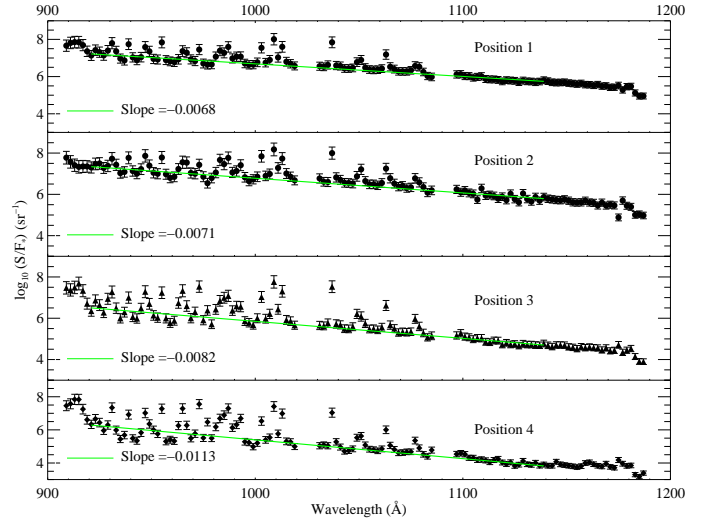}}
\caption{\label{sovf_fuse}The ratio of nebular surface 
brightness to stellar flux at the four positions observed
by $FUSE$ confirm the blue rise discovered by the 
rocket data.  The slope was determined at 
wavelengths where there was no contamination from 
airglow features or interstellar absorption in the Pos1
S/F$_{\star}$.}
\end{center}
\end{figure}

The increased sensitivity and spectral resolution of $FUSE$ allowed us to
confirm the blueness found in IC 405 and to
determine the contribution of fluorescent H$_{2}$ to the nebular surface 
brightness.   HD 34078 has been observed by $FUSE$ under program 
A070 and stellar spectra were obtained from the MultiMission Archive 
at the Space Telescope Science Institute (MAST).  
With the spatial information 
obtained by multiple pointings in the case of the $FUSE$ data, we were 
able to confirm the rocket result independently. 
Figure~\ref{sovf_fuse} shows a similar 
blue S/F$_{\star}$ at all four pointings.  The log$_{10}$(S/F$_{\star}$)
was fit at wavelengths unobstructed by terrestrial airglow lines or
interstellar absorption from 912 -- 1150 \AA\ in the Pos1 spectrum, 
where the scattered stellar continuum was bright
(the long wavelength 
limit was chosen to avoid ``The Worm'' present in the data).  We find 
that the data is well matched by a linear fit 
with a slope of (8.4~$\pm$~0.4)~$\times$~10$^{-3}$, averaged over all positions.
For comparison, the rocket derived S/F$_{\star}$
was fit using the same procedure and the agreement is good,
(8.7~$\pm$~0.6)~$\times$~10$^{-3}$.

\begin{figure}
\begin{center}
\epsscale{0.57}
\rotatebox{90}{
\plotone{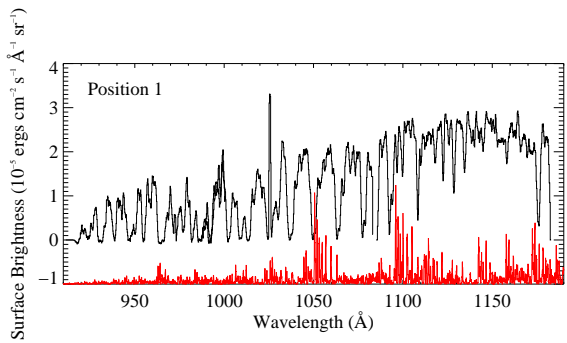} }
\caption{\label{fusespec1}The $FUSE$ spectrum of Pos1. The spectrum is 
dominated by scattered stellar continuum.  The data have 
been smoothed with a 5-pixel boxcar average for display purposes.
The brightest emission feature is geocoronal Ly-$\beta$.  A synthetic 
spectrum of fluorescent H$_{2}$ is overplotted as a red
dashed line to guide the eye.}
\end{center}
\end{figure}

Figures~\ref{fusespec1} and~\ref{fusespec4}  
display the nebular spectra evolving from the influence of
a strong radiation field very near the star 
(Pos1, $\chi$~$\sim$~10$^{5}$, where $\chi$ is the 
average interstellar radiation field;~\citet{draine78}) to the furthest offset, 
approximately 1000\arcsec~to the north (Pos4, $\chi$~$\sim$~10$^{3}$).  Both spectra
are overplotted with a synthetic H$_{2}$ emission spectrum, created by a
fluorescence code similar to the one described in~\citet{wolven97}.
Pos1 and Pos2 are dominated by a 
scattered stellar spectrum, with pronounced H$_{2}$ absorption troughs 
but only hints of H$_{2}$ emission near 
1100 and 1160 \AA.  Pos3 shows the scattered stellar spectrum less strongly 
as the influence of HD 34078 diminishes with distance, and the 
H$_{2}$ emission becomes clear.  The separation from the star is great 
enough at Pos4 ($\approx$~2 pc at the stellar distance of 450 pc) that the 
numerous fluorescent emission lines appear strongly
from 1050 \AA\ to the end of the bandpass near 1187 \AA.  
The observed nebular lines arise from electronic transitions 
($B$$^{1}\Sigma^{+}_{u}$~and~$C$$^{1}\Pi_{u}$~to~$X$$^{1}\Sigma^{+}_{g}$)
that decay from the ground and first excited vibrational state
($\nu$~=~0,1). 

\section{Results and Discussion}

The strong blue rise in IC 405 is remarkable because both observational~\citep{ngc2023,witt93,murthy93}
and theoretical~\citep{wd01} studies find a decreasing albedo 
across the far-UV bandpass.  What mechanism is responsible for 
not only overcoming the falling albedo but increasing
the observed nebular brightness?  We consider:
\begin{itemize}
	\item Peculiar dust grain properties 
	(unusual values of $a$ and $g$) in IC 405,   
	\item Strong fluorescent H$_{2}$ emission that falls 
	below the sensitivity and spectral
	resolution of the rocket experiment,
	\item Unusually small grain distribution leading to a strong 
	Rayleigh scattered component of the nebular brightness,
	\item An unusual dust emission process, an extended blue emission, and
	\item Differential extinction in IC 405 due to 
 	a complicated local geometry or an intervening clump of gas and
	dust along the line of sight to HD 34078 
\end{itemize}

\subsection{Dust Modeling} \label{mod}

%%%%%%%%%%%%%%%%%%%%DUST MODELS%%%%%%%%%%%%%%%%%%%%%%%%
The dust scattering in IC 405 was modeled using a modified version of 
the code described by Burgh et al. (2002), employing a Monte Carlo 
dust radiative transfer model (see also Gordon et al. 2001).\nocite{gordon01}
The model follows the path (direction and position) of each photon in the 
nebula from its ``creation'' at the position of the star, until it leaves the 
nebula (i.e., its radial position is outside the defined size of the nebula).  
The factors that
determine the position and direction of the photons during their propagation 
through the nebula are the optical depth of the dust, the fraction of 
photons scattered by the dust rather than absorbed (the albedo), the 
angular distribution of the scattered photons (parameterized
by $g$ in the scattering distribution given by the Henyey-Greenstein 1941 phase function)\nocite{hg41}, 
and the geometry of the dust distribution.  
Although~\citet{draine03} has argued that the H-G function does not reproduce the
scattering function calculated from the optical properties of dust models
in the far-UV bandpass, we use it here to allow for a direct comparison 
with the result from Burgh et al. (2002). 
A revision of our model replacing the H-G phase function with 
a function that depends on the scattering cross-sections for dust grains in the 
far-UV (as well as more sophisticated geometries) will be addressed in a
future work.
Given inputs for the number of photons followed, the optical depth,
albedo, and $g$, the model outputs an image of the nebular surface brightness
that can be compared to the distribution measured by the rocket experiment.
The rocket data are binned by wavelength region to improve S/N, 
and then plotted as a function of spatial position along the slit.  

\begin{figure}
\begin{center}
\epsscale{0.9}
\rotatebox{90}{
\plotone{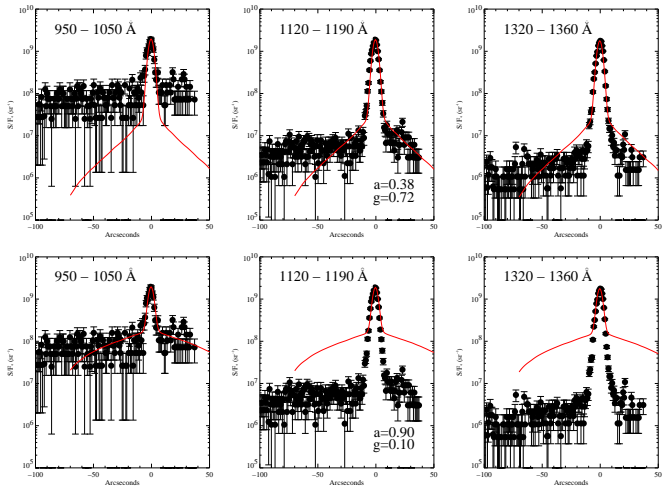} }
\caption{\label{badfit}Spatial S/F$_{\star}$ measured
with the rocket experiment compared with models.
The top panels are models using theoretical values of $a$ and $g$ at 1350 \AA, 
the bottom panels attempt to fit the short wavelength S/F$_{\star}$, using 
a differential extinction of 50$\%$ with $a$=0.9 and $g$=0.1.
The models differ from the one described in Burgh et al. by the use of 
a geometry that employs a larger inner cavity cleared out by the 
stellar wind and high proper motion and includes the differential 
extinction correction.}
\end{center}
\end{figure}

This model, which works well to closely approximate the observed 
surface brightness distribution in the bright reflection nebula
NGC 2023, fails to reproduce the 
nebular surface brightness of IC 405 by orders of magnitude.  Models 
were run for a number of values for the optical depth, $a$, $g$, and
nebular geometry.  Models explored spherical geometries with
constant density and $\rho$~$\propto$~$r^{-\alpha}$, as well as
spherical shell and a constant density slab.
The results are best illustrated by considering two specific 
cases, shown in Figure~\ref{badfit}, the first
using the $a$ and $g$ from Weingartner and Draine
(2001) for R$_{V}$~=~4.0.  The model shows poor agreement across the bandpass, 
particularly at the short wavelength end (a result of the nebular blueness).
The second model finds the best fit to the short
wavelength data, revealing $a$~= 0.9, $g$~= 0.1, in strong disagreement with 
both theoretical and other measured values in this wavelength regime 
(see Draine 2003a and references therein)\nocite{draine03ar}.  
Additionally, a 50$\%$ differential extinction was needed in  
the best-fit 950-1050~\AA\ model; i.e. the stellar flux directly along 
the line of sight to the observer was
reduced by a factor of two.  As this model begins to reproduce the short 
wavelength result, the longer wavelength data is in poor agreement. 
This exercise clearly shows that this reflection nebula is not well
described by a smooth and uniform dust distribution.  More sophisticated
models, employing varying dust densities, complex geometries, and 
more appropriate scattering phase functions than exist at present may
be able to reproduce our observations of IC 405.

\subsection{Comparison with other Reflection Nebulae} \label{comp}

Only one far-ultraviolet observation exists that 
includes both spatial and spectral information,  
however there are a few previous
data sets that put our S/F$_{\star}$ result in context.  
Burgh et al. (2002) reported on observations of NGC 2023 made 
with the same rocket-borne imaging spectrograph described 
above.  This nearly identical observation, made in 2000 February, 
found that the ratio of nebular surface brightness to stellar flux was 
constant with wavelength across the 900 to 1400~\AA\ bandpass.  They 
model the nebular scattering properties (as described above) and 
determine that a decreasing dust albedo is being offset
by grains that are more strongly forward scattering at shorter 
wavelengths (increasing $g$).  

Witt et al. (1993) and Murthy et al. (1993) used observations  
made by the \\
$Hopkins$~$Ultraviolet$~$Telescope$~($HUT$) to 
measure S/F$_{\star}$ in another bright 
reflection nebula, NGC 7023, without the benefit of a spatially resolved slit.  
Murthy et al. describes
the $HUT$ observations made during the Astro-1 mission in 1990
December, obtaining a spectrum of the central star (HD 200775)
and then offsetting to a pointing within the nebula.  They obtained
data with an appreciable signal in the 1100 -- 1860~\AA\ region
and also found that the S/F$_{\star}$ ratio was constant with 
wavelength.  They use a Monte Carlo scattering model 
(described by Witt et al. 1982)\nocite{witt82} and find a 
decreasing albedo for $\lambda$~$<$~1400~\AA\ (assuming $g$~=~0.7).  
This result agrees with that of~\citet{witt93}, who combined
the $HUT$ stellar observation with data from the ultraviolet spectrometer 
on $Voyager$~$2$ and found a drop in the albedo of 25$\%$ between 1300
and 1000~\AA.  The observed S/F$_{\star}$ is found to be flat across this
wavelength region. An increase in nebular surface
brightness of 25$\%$ from an unresolved H$_{2}$ emission component is assumed to
counteract the falling albedo~\citep{witt93,sternberg}.  Ultraviolet studies of a similar
nature have been carried out on the Scorpius OB association~\citep{gordon94}, 
IC 435~\citep{calzetti95}, and the Pleiades reflection nebula~\citep{pleiades03}, 
but these studies focused on longer wavelengths than considered here.

\subsection{Mechanisms for Producing the Blue Rise} \label{dis}
\begin{figure}[b]
\begin{center}
\epsscale{0.57}
\rotatebox{90}{
\plotone{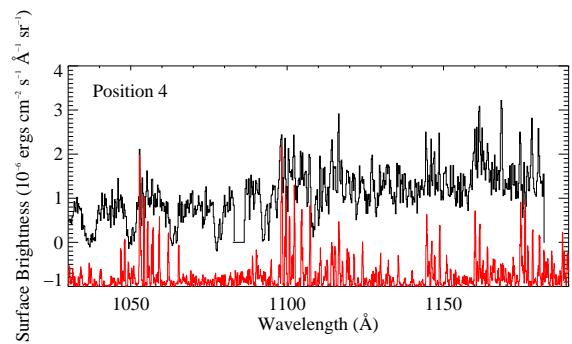} }
\caption{\label{fusespec4}$FUSE$ spectrum of Pos4, 1000\arcsec~north of the star. 
H$_{2}$ emission is now the most prominent spectral feature
from 1050 to 1180~\AA.}
\end{center}
\end{figure}

%%%%%%%%%%%%%%%%%%%%%%%%DISCUSSION%%%%%%%%%%%%%%%%%%%%
As our dust scattering models fail to reproduce the 
short wavelength rise in S/F$_{\star}$ for realistic values of 
grain parameters in IC 405, we consider other possibilities. Ultraviolet 
H$_{2}$ emission is clearly
present in IC 405, the double peaked emission feature near 1600 \AA\ was
first seen by $HUT$ (spectrum ic405\_080) and our $FUSE$ data
resolve the individual rotational components of several vibrational
bands between 1050 and 1185 \AA\ (Figure~\ref{fusespec4}).  Near HD 34078 however, we find  
little evidence for H$_{2}$ emission as either the continuum 
overwhelms the fluorescent signal or the $\chi/n$ environment 
is unfavorable for the fluorescent process~\citep{sternberg}.  Regardless of the relative contribution of 
fluorescing H$_{2}$ to the nebular spectrum, each of our
pointings in IC 405 reveal a very similar, blue S/F$_{\star}$ ratio.  
The evolution of the fluorescence signature with 
distance from the exciting star is interesting, but the constancy of the
nebular brightness to stellar flux rules out molecular hydrogen as the 
dominant cause of the blue rise in IC 405.  

A small grain population has been suggested for NGC 7023~\citep{murthy93,witt93}, 
yet these grains are expected to be efficient absorbers of far-UV 
radiation.  We have no reason to expect to find a unique distribution of 
high albedo grains in IC 405, although we note that the large proper
motion of HD 34078 brings it into contact with grains that are unlikely
to have undergone significant processing by UV photons.  
The exciting stars
of NGC 2023 and 7023 are thought to be born in the presence of the 
dust with which they are interacting, clearly a different scenario than 
in IC 405.  It is interesting to note that~\citet{vijh04} have recently 
found a continuous near-UV/optical
emission in the Red Rectangle nebula which they attribute to fluorescence
by small polycyclic aromatic hydrocarbon (PAH) molecules.  We are unaware
of a molecular continuum process that operates at far-UV wavelengths, however
we cannot conclusively rule out a contribution from PAH molecules.  Further
studies of the vacuum ultraviolet emission and absorption properties
of PAH molecules would be of interest.

\citet{mathis02} have argued that 
clumpy dust distributions  make grain 
properties derived from observations of reflection nebulae unreliable.
They develop a model of hierarchical clumping that predicts widely varying
determinations of the albedo and the scattering parameter
depending on the viewing angle, the degree of clumping, if the central star happens
to reside inside a dense clump, and the density of the interclump medium.
As dust scattering models and molecular hydrogen fluorescence have failed to 
reproduce our data, and other explanations seem speculative at this time,
we adopt differential extinction due to a clumpy dust distribution as 
the most probable explanation for the blue S/F$_{\star}$ ratio in IC 405.  A small
interstellar clump/filamentary structure or circumstellar feature 
directly along the sight line to HD 34078,
but not present in front of the rest of the nebula would 
preferentially extinguish the shorter wavelength stellar photons 
reaching us directly, thus making the denominator in 
S/F$_{\star}$ small at short wavelengths.  Our dust scattering models
are unable to account for such complex geometries, but we can begin to 
study the degree of clumpiness by examining the extinction in IC 405.

\subsection{Differential Extinction Limits}

We can put limits on a differential extinction level that can account
for the blue rise by quantifying the extinction properties and physical
size of an obscuring clump.
We begin by determining the level of
variation in the extinction curve between HD 34078 and the
region surrounding it.  Using the extinction curve to deredden
the HD 34078 line of sight will constrain the 
difference in attenuation between the stellar and nebular lines of sight.  
We deredden the HD 34078 sight line by dividing the stellar spectrum
by the observed extinction curve.  An extinction curve is created
following the pair method, a correction for molecular hydrogen absorption is applied, 
and then the resultant curve is fit using the parameterization
of~\citet{fm90}.  HD 93521, (09V, $E(B~-~V)$~=~0.02; Buss et al. 1995)\nocite{buss95} 
was used as the comparison star and 
the extinction curve was created between 900 and 3000 \AA\ from 
a combination of the rocket data and archival $IUE$ and $HUT$ spectra.
Molecular hydrogen absorption
was corrected for using the $H_{2}ools$ absorption templates for 
the parameters described in $\S$2.2~\citep{h2ools,hurwitz02}.  
We find the FM-fit
parameters to be ($x_{0}$,$\gamma$,$c_{1}$,$c_{2}$,$c_{3}$,$c_{4}$)~=~(4.59,0.94,0.17,0.37,6.25,1.05).
%x0,gam,c1,c2,c3,c4 = 4.589,0.937,0.173,0.372,6.251,1.050
Repeating our S/F$_{\star}$ analysis using the dereddened star, 
we find that the ratio is constant across the rocket bandpass.
Assuming that the extinction curve only applies to the stellar sight line seems
to account for the
blue rise, but comes with the implication that
the nebula has a foreground extinction of zero.  If differential 
extinction is the sole mechanism at work in creating a blue S/F$_{\star}$
in IC 405, then all of the observed reddening would appear to originate in a small clump
along a pencil-beam to HD 34078.

We determine the size of such a ``small clump'' 
by measuring the nebular brightness profile along the  rocket slit at 
the ``on-star'' pointing, and find the the clump size must be smaller
than about 20\arcsec.  It is unclear where the clump is located along the
line of sight to HD 34078, but an upper limit on the physical size can be 
set by requiring it to be near the star, we find the upper limit to be 
0.04 pc.  Variations have recently been detected 
in the column density of CH along the line of sight to HD 34078, 
suggesting structure on the scale of tens of AU~\citep{rollinde03}.  One 
possible explanation for the $N$(CH) variation is the presence of 
a clump along the HD 34078 line of sight, newly present due to 
the high proper motion of the star, although the the analysis presented in
\cite{rollinde03} finds this hypothesis to be unlikely.  
Independent of the location of the clump, there is other evidence of multiple
absorption components along the line of sight.
In addition to the scattered stellar continuum seen at Pos1, $FUSE$ spectra 
of HD 34078 revealed both cold (T~=~80 K) and highly 
excited H$_{2}$ in absorption~\citep{fuse2,rollinde03}. 
The ambient interstellar radiation field 
is insufficient to excite H$_{2}$ to the observed levels 
(up to $\nu$~=~0, j~=~11), implying the relative proximity of HD 34078.

\subsection{Consequences for Unresolved Observations}

We note that if IC 405 were observed in a spatially unresolved manner or 
seen from a large distance, HD 34078 would appear to be less extinguished than
in the present case, where the star and nebula can be measured separately.
If observed at a large distance, the 
integrated blue nebular continuum and the reddened star light would be
considered stellar in origin, artificially dereddening the intrinsic stellar continuum.  
As a rough determination of how such an observation may be biased, 
we assume a uniform surface brightness
filling the rocket slit at the on-star pointing, and estimate the contribution of the 
nebular flux to the total observed spectrum in the unresolved case.  In the 
simple case of a constant surface brightness nebula the length of the rocket
long-slit on each side (200\arcsec$\times$200\arcsec,~$\approx$~0.4 pc$\times$0.4 pc), 
we determine a 
nebular filling factor correction of about 15.  Multiplying the nebular flux
measured during the on-star position by the filling factor and adding that to the 
stellar flux, we 
determine how bright HD 34078 would appear in an unresolved observation.  Taking
a constant flux of $\approx$~1.0$\times$~10$^{-12}$~ergs~cm$^{-2}$~s$^{-1}$~\AA$^{-1}$
at 1100 \AA\ (from the nebular spectrum observed by the rocket, consistent 
with the $FUSE$ measurement), we find that the integrated nebular flux would be
$\approx$~1.5$\times$~10$^{-11}$~ergs~cm$^{-2}$~s$^{-1}$~\AA$^{-1}$, 50$\%$ brighter
than the star itself.  A spatially unresolved far-UV observation of IC 405 would 
lead one to measure a stellar flux level 2.5 times (one magnitude) greater than the true value.
This effect is reminiscent of
the observed UV properties of local starburst galaxies described by~\citet{calzetti97}.
Local starbursts show less extinction of the stellar continuum relative
to the nebular emission lines, explained by a clumpy dust distribution 
that preferentially reddens nebular light due to the different filling
factors of the stars and gas.  Calzetti (1997) finds that stellar photons
only encounter approximately 60$\%$ of the dust seen by nebular photons.

\section{Summary}

The emission/reflection nebula IC 405 was observed by a rocket-borne, long-slit imaging
spectrograph in the far-ultraviolet bandpass (900 -- 1400 \AA).  
A high quality spectrum (S/N~$\approx$~10 -- 15) 
was obtained of the central star, HD 34078, as 
well as the spectra of the surrounding nebula out to an offset 
approximately a parsec from the star.  We found that the ratio 
of nebular surface brightness to stellar flux (S/F$_{\star}$) rose by two 
orders of magnitude to the blue across the bandpass of the instrument.
This result is in conflict with analogous observations of reflection
nebulae in this wavelength regime (NGC 2023 and 7023), where 
flat S/F$_{\star}$s were observed.  This result 
held true for both positions within the nebula.  Additional observations
were made with $FUSE$ in an attempt to clarify the process responsible 
for the blue rise.  $FUSE$ observed four positions within IC 405, 
three coincident with the rocket pointings, and one along another bright
nebular filament.  These data revealed the nature of the nebular 
spectra: a progression from strong scattered stellar continuum near the star to 
the appearance of fluorescent emission from H$_{2}$ farthest from the star.  
An analysis of S/F$_{\star}$ confirmed the blueness
throughout IC 405, showing little correlation with the nebular 
spectral characteristics.

Models of scattering in a uniform dust distribution,  
similar to those that accurately reproduce
the observed  S/F$_{\star}$ in NGC 2023, were unable to fit 
IC 405 for  realistic values of the albedo and the phase function
asymmetry parameter for the geometries studied.  The presence
of fluorescent H$_{2}$ creating excess emission to the blue 
while going undetected at the resolution of the rocket experiment
has been ruled out by $FUSE$, as S/F$_{\star}$ remains constant regardless
of the relative contribution of H$_{2}$ to the spectrum.  More exotic
explanations, such as an extended blue emission from dust, are speculative. 
We favor the hypothesis of differential extinction as caused by a 
clumpy dust distribution, which has been suggested by Mathis et al. (2002)
to complicate the conclusions drawn from observations of reflection nebulae.
Differential extinction along the line of sight to IC 405, such as a knot or
filament of gas and dust on the sight line to HD 34078 but not 
crossing the path to the rest of the nebula can account for the blue rise
if one assumes the obscuring material is very local to the HD 34078
line of sight.  Our observations have 
placed an upper limit to the size of the intervening clump of 0.04 pc.  
If clumpy dust is responsible for the blue rise that we have found, 
it seems that IC 405 is the prototypical example for arguments
against the use of reflection nebulae for reliable determination of dust
grain properties.  IC 405 appears to be a local example of the 
differential extinction process that takes place in starburst systems on a
global scale.

\acknowledgments

We thank~\nocite{draine03,gordon01,srm94,moos00,sanhow00,mathis02,ccm,witt82}
Russell Pelton of the JHU Sounding Rocket Group for his 
dedication to every phase of the 36.198~UG mission.  We wish to 
acknowledge NASA's Wallops Flight 
Facility personnel, the support 
staff at White Sands Missile Range and the Physical Sciences Laboratory, 
operated by New Mexico State University, for their professional support. 
We also wish to thank Brad Frey for his assistance with 
calibration and field operations.
The rocket data was 
supported by NASA grant NAG5-5122 to the Johns Hopkins University and the 
$FUSE$ data was obtained under the Guest Investigator Program by the 
NASA-CNES-CSA $FUSE$ mission, operated by the Johns Hopkins University.

%%%%%%%%%%%%%%%%%%%%%%%%%%%BIBLIOGRAPHY%%%%%%%%%%%%%%%%%%

\bibliography{ic405pap}


\end{document}